\documentclass[]{article}
\usepackage{stmaryrd,amssymb,amsmath,amsthm,url}
\usepackage[all]{xy}

\title{Tuplix Calculus\thanks{%
	Thanks are due to Sanne Nolst Trenit\'e 
	(UvA, Faculty of Science) for many discussions
	and for helping to clarify the usefulness and necessity
	of formalization in budget calculations.}
}
\author{
	Jan A.\ Bergstra \and
	Alban Ponse \and
	Mark B.\ van der Zwaag\\
\\
  {\small
	  Section Software Engineering,
	  Informatics Institute,
	  University of Amsterdam}\\
	{\small Email: \url{{janb,alban,mbz}@science.uva.nl}
	}
}
\date{}

\newtheorem{theorem}{Theorem}
\newtheorem{lemma}{Lemma}
\theoremstyle{definition}
\newtheorem*{example}{Example}
\newtheorem*{note}{Note}

\newcommand{\axname}[1]{\ensuremath{\textup{#1}}}
\newcommand{\defeq}{\ensuremath{\stackrel{\text{\tiny{def}}}{=}}}
\newcommand{\CTC}{\ensuremath{\axname{CTC}}}
\newcommand{\BTC}{\ensuremath{\axname{BTC}}}
\newcommand{\SUBST}{\ensuremath{\axname{T10}^{{+}}}}
\newcommand{\DE}{\ensuremath{\axname{\textsc{De}}}}
\newcommand{\DEP}{\ensuremath{\DE^{{+}}}}
\newcommand{\Attr}{\ensuremath{\mathit{A}}}
\newcommand{\et}{\ensuremath{\varepsilon}} 
\newcommand{\nt}{\ensuremath{\delta}}
\newcommand{\conjc}{\ensuremath{\varobar}}
\newcommand{\ztest}[1]{\ensuremath{\gamma(#1)}}
\newcommand{\nztest}[1]{\ensuremath{\widetilde{\gamma}(#1)}}
\newcommand{\FV}{\ensuremath{\mathit{FV}}}
\newcommand{\V}{\ensuremath{\mathit{Var}}}
\newcommand{\gsum}{\ensuremath{\textstyle{\sum}}}
\newcommand{\Clr}{\ensuremath{\et}}
\newcommand{\Enc}{\ensuremath{\partial}}
\newcommand{\SM}{\ensuremath{\mathcal{M}}}
\newcommand{\DM}{\ensuremath{\mathcal{D}}}
\newcommand{\sem}[1]{\ensuremath{\llbracket#1\rrbracket}}
\newcommand{\PF}{\ensuremath{F}}
\newcommand{\Budget}{\ensuremath{B}}
\newcommand{\perA}{\ensuremath{\mathrm{A}}}
\newcommand{\perB}{\ensuremath{\mathrm{B}}}
\newcommand{\source}{\ensuremath{\mathrm{S}}}
\newcommand{\control}{\ensuremath{\mathrm{C}}}
\newcommand{\pu}{\ensuremath{\mathrm{PU}}}
\newcommand{\pua}{\ensuremath{\pu_{1}}}
\newcommand{\pub}{\ensuremath{\pu_{2}}}
\newcommand{\service}{\ensuremath{\mathrm{SC}}}
\newcommand{\capa}{\ensuremath{\mathrm{CAP}}}
\newcommand{\reward}{\ensuremath{\mathit{reward}}}
\newcommand{\Rone}{\ensuremath{\reward_{1}}}
\newcommand{\Rtwo}{\ensuremath{\reward_{2}}}
\newcommand{\nop}{\ensuremath{\mathit{n}}}
\newcommand{\sfrac}{\ensuremath{\mathit{k}}}
\newcommand{\attrname}[1]{\ensuremath{\mathit{#1}}}

\begin{document}

\maketitle

\begin{abstract}
We introduce a calculus for tuplices,
which are expressions that generalize matrices and vectors.
Tuplices have an underlying data type for \emph{quantities}
that are taken from a zero-totalized field.
We start with the core tuplix calculus \CTC\
for entries and tests,
which are combined using conjunctive composition.
We define a standard model and prove
that \CTC\ is relatively complete with respect to it.
The core calculus is extended with operators for
choice, information hiding,
scalar multiplication, clearing and encapsulation.
We provide two examples of applications;
one on incremental financial budgeting,
and one on modular financial budget design.
\end{abstract}

\section{Introduction}
\label{sec:Intro}
In this paper we propose \emph{tuplix calculus}:
a calculus for so-called \emph{tuplices},
which are expressions that generalize matrices and vectors.
Tuplices have an underlying data type called \emph{quantities}.
We shall require that this data type is modeled
by a zero-totalized field, 
in the terminology of~\cite{BT07,BHT07}, 
as will be explained in Section~\ref{sec:ZTF}. 
A typical example of a tuplix is a budget, a compound of 
various attributes, each of which possesses a certain value
(a quantity) and may refer to certain conditions and/or
interdependencies.
Another example of a tuplix is the modeling
of \emph{let\/}-expressions in functional programming. 
We provide a standard model for tuplix calculus and 
discuss some examples of its use.

A tuplix generalizes a vector or a matrix in that it
collects a number of quantities from the same data type
under a number of names (dimensions of the vector,
matrix entries). 
What differs in the design of tuplix calculus from matrix or 
vector calculus is that other methods of compositional 
construction are envisaged.
\emph{Conjunctive composition} extends both vector (matrix) 
addition and set union into a novel compositional mechanism. 
In addition, `information hiding' is provided which supports
the use of auxiliary values whose name is made externally
invisible. 
In order to provide a simple semantic model of this form
of information hiding, \emph{alternative composition} of
tuplices is included as well.
Alternative composition, written $x+y$ 
denotes a tuplix which is either $x$ or $y$.
Information hiding is modeled using
\emph{generalized alternative composition}.
A further key feature of tuplix calculus 
is the inclusion of conditions as atomic tuplices;
this is done via a \emph{zero-test operator}.  
Finally an \emph{encapsulation} mechanism is proposed. 
Encapsulation removes an entry and enforces that it
will contain quantity zero only.
Encapsulation and alternative composition 
work together exactly as in the process
algebra ACP \cite{BK84} from which the typescript has
been borrowed
(see~\cite{BW90} and~\cite{F00} for more recent expositions of
ACP-style process algebra).

Tuplices constitute a calculus rather than an algebra
because information hiding introduces bound variables.
The motivation for designing tuplix calculus came from
a number of attempts to design financial budgets in a
modular fashion.
One might call a tuplix a budget but we prefer not to
introduce that financial connotation by using a mathematically
neutral term which can be viewed as describing a purely
structural notion without any preferred application or
even application area.
We call tuplix calculus an abstract data type calculus.
It is based on an algebraic abstract data type but this is 
augmented with operators involving bound variables,
notably the generalized alternative composition operator.

The design of tuplix calculus is based on zero-totalized
fields because this drastically simplifies type checking
in general and equational logic in particular for fields. 
That zero-totalized fields are meadows, which in general
may feature proper zero-divisors as a natural generalization,
is of less importance to the design of tuplix calculus.

The paper is structured as follows.
Section~\ref{sec:ZTF} discusses zero-totalized fields.
Section~\ref{sec:CTC} introduces the axiom system \CTC\
(Core Tuplix Calculus).
We define a standard model for the interpretation
of tuplices and prove the relative completeness of \CTC\
(relative: valid data identities are assumed
in the proof theory) with respect to this standard model
(Section~\ref{sec:SM}).
In the next sections, \CTC\ is extended with several
operators, starting with alternative composition
in Section~\ref{sec:BTC},
leading to Basic Tuplix Calculus (\BTC).
Section~\ref{sec:ZTL} is an intermezzo containing some
observations on the use of zero tests (\emph{zero-test logic}).
Section~\ref{sec:gensum} introduces information hiding
to \BTC\ through the binding of data variables.
Section~\ref{sec:aux} defines three auxiliary operations
including encapsulation.
Sections~\ref{sec:vb1} and~\ref{sec:vb2} present example
applications.
We end with some concluding remarks (Section~\ref{sec:concl}).

\section{Cancellation Meadows}\label{sec:ZTF}
Quantities will be taken from a 
\emph{non-trivial cancellation meadow}, or,
equivalently, from a \emph{zero-totalized field},
in the terminology of \cite{BT07,BHT07}. 
A zero-totalized field is 
the well-known algebraic structure `field'
with a total operator for division so 
that the result of division 
by zero is zero
(and, for example, in a 47-totalized field one has chosen 47 to represent 
the result of all divisions by zero). 

A \emph{meadow} is a commutative ring with unit equipped with
a total unary operation $(\_)^{-1}$
named inverse that satisfies the axioms
\[
  (u^{-1})^{-1} = u   \quad\text{and}\quad
  u\cdot(u \cdot u^{-1}) = u,
\]
and in which $0^{-1}=0$.
For quantities (and tuplix calculus) we also
require the 
\emph{cancellation axiom}
\[
  u\neq 0 \quad \& \quad u\cdot v=u\cdot w\quad\Rightarrow\quad v=w
\]
to hold,
thus obtaining \emph{cancellation meadows}, which we take as
the mathematical structure for quantities, requiring further
that $0\neq 1$ to exclude (trivial) one-point models.
These axioms for cancellation meadows
characterize exactly
the equational theory of zero-totalized fields~\cite{BHT07}.
The property of cancellation meadows that is
exploited in the tuplix calculus is
that division by zero yields zero, 
while $u\cdot u^{-1}=1$ for $u\neq 0$.

For the tuplix calculus,
we define a
\emph{data type} (signature and axioms) for quantities
which comprises the constants 
0, 1, the binary operators $+$ and $\cdot$,
and the unary operators
$-$ and $(\_)^{-1}$.
We often write
$u-v$ instead of $u+(-v)$,
$u/v$ instead of $u\cdot v^{-1}$, and
$uv$  instead of $u\cdot v$,
and we shall omit brackets if no confusion can arise
following the usual binding conventions.
Finally, we use numerals in the common way
(2 abbreviates $1+1$, etc.).
The axiomatization 
consists of the cancellation axiom
\[
  u\neq 0 \quad \& \quad u\cdot v=u\cdot w\quad\Rightarrow\quad v=w,
\]
the \emph{separation axiom} 
\[
  0\neq 1,
\]
and the following 10 axioms for meadows (see \cite{BHT07}):
\begin{align*}
	(u+v)+w &= u + (v + w),\\
	u+v     &= v + u .\\
	u+0     &= u ,\\
	u+(-u)  &= 0 ,\\\displaybreak[0]
	(u \cdot v) \cdot  w &= u \cdot  (v \cdot  w),\\
	u \cdot  v &= v \cdot  u,\\
	1\cdot u &= u, \\\displaybreak[0]
	u \cdot  (v + w) &= u \cdot  v + u \cdot  w ,\\
	(u^{-1})^{-1} &= u ,\\
	u \cdot (u \cdot u^{-1}) &= u.
\end{align*}

The following identities are derivable from the axioms
for meadows.
\begin{align*}
	(0)^{-1}  &= 0\\
	(-u)^{-1} &= -(u^{-1})\\
	(u \cdot  v)^{-1} &= u^{-1} \cdot  v^{-1}\\
	0\cdot u  &= 0\\
	u\cdot -v &= -(u\cdot v)\\
	-(-u)     &= u
\end{align*}
Furthermore,
the cancellation axiom
and axiom $u \cdot (u \cdot u^{-1}) = u$ 
imply the \emph{general inverse law} 
\[
  u\neq 0 \quad\Rightarrow\quad u\cdot u^{-1}=1
\]
of zero-totalized fields.

\section{Core Tuplix Calculus}
\label{sec:CTC}
Tuplix calculus builds on the
data type defined in Section~\ref{sec:ZTF}
which specifies non-trivial cancellation meadows.
We use the letters $u$, $v$ and $w$ as data variables,
and the letters $p$ and $q$ to range
over (open) data terms.

We start with a core calculus
which can be extended with several
operators (as is done in later sections).
The theory is parametrized with a nonempty set \Attr\
of \emph{attributes}, ranged over by $a$ and $b$.
We further assume given a countably-enumerable
set of tuplix variables, ranged over
by $x$, $y$ and $z$.
We introduce the signature for tuplices.
We have constants \et\ (the empty tuplix)
and \nt\ (the null tuplix);
the variables are tuplix terms; and
there are two further kinds of atomic tuplices:
\emph{entries} (attribute-value pairs) of the form
\[ a(p)
\]
with $a\in\Attr$, and $p$ a data term, 
and, for any data term $p$, the \emph{zero test}
\[ \ztest{p}
\]
($\gamma\not\in\Attr$).
Finally, the core theory has one binary infix operator:
the \emph{conjunctive composition} operator \conjc.
This operator is commutative and associative.

Axioms:
\begin{align}
\tag{\axname{T1}}\label{ax:T1} x\conjc y &= y\conjc x\\
\tag{\axname{T2}}\label{ax:T2} 
	(x\conjc y)\conjc z &= x\conjc(y \conjc z)\\
\tag{\axname{T3}}\label{ax:T3} x\conjc\et &= x\\
\tag{\axname{T4}}\label{ax:T4} x\conjc\nt &= \nt\\\displaybreak[0]
\tag{\axname{T5}}\label{ax:T5} a(u)\conjc a(v) &= a(u+v)\\
\tag{\axname{T6}}\label{ax:T6}\ztest u &= \ztest{u/u}\\
\tag{\axname{T7}}\label{ax:T7}\ztest 0 &= \et\\
\tag{\axname{T8}}\label{ax:T8}\ztest 1 &= \nt
\end{align}

In this core calculus, a tuplix is 
a conjunctive composition of tests and entries,
with \et\ representing an empty tuplix,
and \nt\ representing an erroneous situation
which nullifies the entire composition.
Entries with the same attribute can be combined
to a single entry containing the sum of the quantities
involved (axiom \ref{ax:T5}).

A zero test \ztest{p} acts as a conditional:
if the argument $p$ equals zero,
then the test is void and disappears
from conjunctive compositions.
If the argument is not equal to zero,
the test nullifies
any conjunctive composition containing it.
Observe how we exploit the property
of zero-totalized fields that $p/p$ is
always defined, and that the division $p/p$ yields 
zero if $p$ equals zero, and 1 otherwise.
Further observe that an equality test $p=q$ can be
expressed as \ztest{p-q}.

A tuplix term is \emph{closed}
if it is does not contain tuplix variables and
also does not contain data variables.
A tuplix term is \emph{tuplix-closed}
if it does not contain tuplix variables
(but it may contain data variables).
For reasoning about tuplices with open data terms, 
we add the following two axioms:
\begin{align}
\tag{\axname{T9}}\label{ax:T9}
	\ztest{u}\conjc\ztest{v}&= \ztest{u/u+v/v}\\
\tag{\axname{T10}}\label{ax:T10}
	\ztest{u-v}\conjc a(u)&= \ztest{u-v}\conjc a(v)
\end{align} 

The tuplix calculus is two-sorted.
On the tuplix side we have the axioms 
\ref{ax:T1}--\ref{ax:T10}
and we use the proof rules of equational logic.
On the data side, we refrain from
giving a precise proof theory.
We adopt the following rule
to lift the valid data identities
to the tuplix calculus:
for all (open) data terms $p$ and $q$,
\begin{equation}\tag{\DE}\label{ax:DE}
  \DM\models p = q
  \quad\text{implies}\quad
  \ztest p =\ztest q, 
\end{equation}
where \DM\ (a non-trivial cancellation meadow)
is our model of the data type.
This axiom system with axioms
\ref{ax:T1}--\ref{ax:T10} plus proof rule~\ref{ax:DE}
is denoted by \CTC\ (Core Tuplix Calculus).

\section{Canonical Terms and Derived Proof Rules}
A \CTC\ canonical term is a term of the form
\[ 
  \ztest{p_0}\conjc 
  a_1(p_1)\conjc\cdots\conjc a_k(p_k)\conjc
  x_{1}\conjc\cdots\conjc x_{l},
\]
for some $k,l\geq 0$,
and with distinct attributes $a_i$ for $i=1,\ldots,k$.

\begin{lemma}\label{lem:bt}
	Every \CTC\ term is derivably equal to
	a \CTC\ canonical term.
\end{lemma}

\begin{proof}
Easy: If it contains the constant \nt,
the term equals the canonical term
\ztest{1} (using axioms \ref{ax:T4} and \ref{ax:T8});
conjunctive composition is commutative and associative;
the \et\ constant disappears (axiom \ref{ax:T3});
entries with same attribute are combined
using axiom \ref{ax:T5}; tests are combined
using axiom \ref{ax:T9} (and if there are no tests, we add a 
void \ztest{0} test using axiom \ref{ax:T7}).
\end{proof}

The axiom system \CTC\ is powerful enough for
our purposes, as is witnessed by
the completeness result in Section~\ref{sec:SM}.
Still, more general proof rules for the derivation of
identities involving data equalities and
substitution of data terms can be convenient.
For example, we find that \CTC\ derives
the ``obvious'' identity
\[
   a(u+v) = a(v+u)
\]
rather indirectly:
because $(u+v)-(v+u)=0$ will 
be valid in our data model, we have
\[
  \ztest{(u+v)-(v+u)}=\ztest 0
\]
by \ref{ax:DE}. Then we derive
\[
  a(u+v) = a(u+v)\conjc \ztest{(u+v)-(v+u)} = a(v+u)
\]
using axioms \ref{ax:T3}, \ref{ax:T7}, and \ref{ax:T10}.

The following proof rule generalizes \DE:
\begin{equation}
\tag{\DEP}\label{ax:DEP}
  \DM\models p = q
  \quad\text{implies}\quad
  t[p/u] = t[q/u],
\end{equation}
for tuplix terms $t$ and
with substitution $t[p/u]$ defined as usual
for two-sorted equational logic
(replacement of all data variables $u$ in $t$ by $p$).
The following axiom scheme generalizes
axiom~\ref{ax:T10}:
\begin{equation}
\tag{\SUBST}\label{ax:subst}
t\conjc\ztest{u-p} = t[p/u]\conjc\ztest{u-p},
\end{equation}
where $t$ ranges over tuplix terms.

These two rules follow from \CTC\ 
as we shall now prove.
We start with two lemmas.

\begin{lemma}\label{lem:lemA}
	For all data terms $p$ and $q$,
	\begin{equation*}
	 \DM\models (1-p/p)\cdot q= 0
	 \quad\text{implies}\quad
	 \CTC\vdash\ztest{p}=\ztest{p}\conjc\ztest{q}.
	\end{equation*}
\end{lemma}

\begin{proof}
Assume that $\DM\models (1-p/p)\cdot q=0$.
Observe that it follows that
\begin{equation*}
 p/p=(p/p+q/q)/(p/p+q/q)
\end{equation*}
is a valid identity
(check: distinguish cases $p=0$ and $p\neq 0$).
From this, derive
\begin{align*}
\ztest p &= \ztest{p/p} \\
  &= \ztest{(p/p+q/q)/(p/p+q/q)}\\
  &= \ztest{p}\conjc\ztest{q}
\end{align*}
using \ref{ax:DE}
and axioms \ref{ax:T6} and \ref{ax:T9}.
\end{proof}

Note that a test \ztest{(1-p/p)\cdot q} may be read
as the logical implication `$p= 0$ implies $q=0$',
see also Section~\ref{sec:ZTL}. 

\begin{lemma}
\label{lem:lemB}
The following identity is derivable in \CTC.
\[
	\ztest u\conjc\ztest{u-v} =
	\ztest v\conjc\ztest{u-v} 
\]
\end{lemma}

\begin{proof}
Observe that
\[
  \left(1- \frac{u/u +(u-v)/(u-v)}{u/u +(u-v)/(u-v)}\right)\cdot v =0
\]
(`if $u=0$ and $u=v$, then $v=0$')
is valid in any cancellation meadow.
Derive
\begin{align*}
\ztest u\conjc\ztest{u-v} 
	&= \ztest{u/u+(u-v)/(u-v)}\\
	&= \ztest{u/u+(u-v)/(u-v)}\conjc\ztest v\\
	&= \ztest u\conjc\ztest{u-v}\conjc\ztest v
\end{align*}
using Lemma~\ref{lem:lemA} and axiom~\ref{ax:T9}.
The remaining part of the derivation is symmetrical.
\end{proof}

We are now ready to derive the two rules.

\begin{itemize}
\item
Case \DEP.
Assume that $\DM\models p=q$, 
and let $t$ be a canonical term
\[
  \ztest{p_0}\conjc a_1(p_1)\conjc\cdots\conjc a_k(p_k)\conjc
  x_1\conjc\cdots\conjc x_l,
\]
for some $k,l\geq 0$.
First observe that it follows from $\DM\models p=q$
that
\[
  \DM\models p_i[p/u]=p_i[q/u]\quad\text{and}\quad
  \DM\models p_i[p/u]-p_i[q/u]=0
\]
for $i=0,\ldots,k$.
From this and \ref{ax:DE}
we derive that
\[
  \ztest{p_0[p/u]} = 
  \ztest{p_0[q/u]}
\]
and
\begin{align*}
  a_i(p_i[p/u]) &= a_i(p_i[p/u])\conjc\ztest 0\\
  &= a_i(p_i)[p/u]\conjc\ztest{p_i[p/u]-p_i[q/u]},
\end{align*}
so we can apply the required substitutions
in the entries using axiom \ref{ax:T10}.

\item Case \ref{ax:subst}.
Let $t$ be a canonical term
\[
  \ztest{p_0}\conjc a_1(p_1)\conjc\cdots\conjc a_k(p_k)\conjc
  x_1\conjc\cdots\conjc x_l,
\]
for some $k,l\geq 0$.
Observe that for $i=0,\ldots,k$,
\[
  \DM\models (1-(u-p)/(u-p))\cdot (p_i-p_i[p/u]).
\]
Therefore we have by Lemma~\ref{lem:lemA} that
\[
  \ztest{u-p} = \ztest{u-p}\conjc\ztest{p_i-p_i[p/u]}
\]
so that we can perform the
substitutions in the entries using axiom~\ref{ax:T10},
and in the test $\ztest{p_0}$ using
Lemma~\ref{lem:lemB}.
\end{itemize}

\section{Standard Model and Relative Completeness}
\label{sec:SM}
We interpret tuplix terms in
the \emph{standard model} $\SM(\DM,\Attr)$,
where \DM\ is the model of
the data type for quantities, and
\Attr\ is the set of attributes that are used. 
The data model \DM\ is required to be a
non-trivial cancellation meadow.
We write \DM\ for the domain of \DM, and $0$
for the element of \DM\ that is the interpretation
of the data term $0$.

The standard model is based on the set
\[
  \PF = \Attr\stackrel p\rightarrow\DM
\]
of partial functions from \Attr\ to \DM\
which are used to model the entries (the attribute-value pairs).
The domain for the standard model is the power set
\[ 
  2^\PF
\]
of the set of partial functions.
An element of this power set stands for a number of
\emph{alternatives}:
for \CTC, the interpretation of tuplix terms yields
either the empty set
(the interpretation of \nt; absence of alternatives)
or a singleton set.
When we add \emph{choice} to the theory
(see Section~\ref{sec:BTC}),
the interpretation may yield sets with more than one element.

Some preliminaries:
\begin{enumerate}
\item For $a\in\Attr$, $d\in\DM$, let $f_{a,d}$ 
be the partial function with $f_{a,d}(a)=d$,
and $f_{a,d}(b)$ undefined for $b\neq a$.

\item We denote by $f_\et$ the 
function in \PF\ with
$f_\et(a)$ undefined for all attributes $a\in\Attr$;
this function will be used in the interpretation 
of the term \et.

\item Define conjunctive composition ${\conjc}$
on elements of \PF\ as follows:
for $a\in\Attr$,
if both $f$ and $g$ are undefined for $a$,
 then $(f\conjc g)$ is undefined for $a$;
if $f(a)$ is defined and $g$ is not defined for $a$,
 then $(f\conjc g)(a) = (g\conjc f)(a) = f(a)$;
and if both $f$ and $g$ are defined for $a$, then
 $(f\conjc g)(a) = f(a)+g(a)$.

\end{enumerate}

The closed terms of \CTC\
are interpreted in the standard model
as follows.
\begin{align*}
\sem\nt &\defeq \emptyset\\
\sem\et &\defeq \{f_\et\}\\
\sem{a(p)} &\defeq \{f_{a,\sem p}\}\\
\sem{\ztest p} &\defeq 
  \begin{cases}
    \sem\et & \text{if } \sem p = 0 \\
	\sem\nt & \text{otherwise}
  \end{cases}\\
\sem{s\conjc t} &\defeq 
  \{ f\conjc g~|~f\in\sem s,\ g\in\sem t \}
\end{align*}

We say that closed terms $s$ and $t$ are \emph{equivalent}
with respect to the standard model if $\sem s=\sem t$.
Two open terms are equivalent,
notation $s\sim t$, if all their closed 
instantiations are pair-wise equivalent.
The axiom system \CTC\ is sound with respect to
the standard model, i.e.,
for all (open) tuplix terms $s$ and $t$,
$\CTC\vdash s=t$ implies $s\sim t$.

\begin{theorem}
  \label{thm:ctccomp}
	The axiom system \CTC\ is complete
	with respect to the standard model, i.e.,
	for all (open) terms $s$ and $t$,
	$s\sim t$ implies $\CTC\vdash s=t$.
\end{theorem}

This completeness is \emph{relative}
in the sense that our proof theory assumes,
by adoption of rule \ref{ax:DE},
all valid data identities.

\begin{proof}
Suppose $s\sim t$.
Using Lemma~\ref{lem:bt} we know 
that $s$ and $t$ are derivably equal to canonical terms
\[
  s'=\ztest{p_0}\conjc a_1(p_1)\conjc\cdots\conjc a_k(p_k)\conjc
     x_1\conjc\cdots\conjc x_l
\]
and
\[
  t'=\ztest{q_0}\conjc a_1(q_1)\conjc\cdots\conjc a_k(q_k)\conjc
     x_1\conjc\cdots\conjc x_l
\]
with $k,l\geq 0$.
Observe that it follows
from $s\sim t$ that we can find canonical terms 
having the same tuplix variables $x_i$ and
(mutually distinct) attributes $a_j$.

It also follows from $s\sim t$, that
whenever the test \ztest{p_0} succeeds,
also the test \ztest{q_0} succeeds, and vice versa.
Therefore, the cancellation meadow identity
$p_0/p_0 = q_0/q_0$ must be valid.
It follows that 
\[
  \ztest{p_0}=\ztest{q_0}
\]
is derivable using axiom \ref{ax:T6} and~\ref{ax:DE}.

It further follows from $s\sim t$, that
whenever the test \ztest{p_0}, and hence
also \ztest{q_0}, succeeds, then it must be 
that $p_i=q_i$ for $i=1,\ldots,k$.
A consequence is that the cancellation meadow identity
\[
  (1-p_0/p_0)(p_i-q_i)=0
\]
is valid
(check: straightforward case distinction on $p_0$).
Using Lemma~\ref{lem:lemA} we find that
\[
  \ztest{p_0} = \ztest{p_0}\conjc\ztest{p_i-q_i}.
\]
Because we also have $\ztest{p_0}=\ztest{q_0}$
it is easy to see that $s'=t'$ is derivable
using axiom \ref{ax:T10}.
\end{proof}

\section{Basic Tuplix Calculus}
\label{sec:BTC}
The axiom system \CTC\ is extended to
{Basic Tuplix Calculus} (\BTC),
by addition of the binary operator $+$ called 
\emph{alternative composition}
or \emph{choice} to the signature,
and by adoption of the following axioms.
\begin{align}
\tag{\axname{C1}}\label{ax:C1} x+y    &= y+x\\
\tag{\axname{C2}}\label{ax:C2}(x+y)+z &= x+(y+z)\\
\tag{\axname{C3}}\label{ax:C3} x+x    &= x\\
\tag{\axname{C4}}\label{ax:C4} x+\nt  &= x\\
\tag{\axname{C5}}\label{ax:C5}
	x\conjc (y+z) &= (x\conjc y)+(x\conjc z)\\
\tag{\axname{C6}}\label{ax:C6}
	\ztest{u}+\ztest{v} &= \ztest{uv}
\end{align}
Because choice is an associative operator, we shall often omit
brackets in repeated applications.

The standard model $\SM(\DM,\Attr)$
for \CTC\ is extended to \BTC\ by
the following interpretation
of alternative composition:
\[
  \sem{s+t} \defeq \sem{s}\cup\sem{t}.
\]
So, the interpretation of
a closed term yields a set of
\emph{alternatives}.
Note that \nt\ is a zero element for alternative composition:
it stands for the absence of alternatives
(recall that $\sem\nt=\emptyset$).
The axioms \ref{ax:C1}--\ref{ax:C6}
are sound with respect to the standard model.

As for \CTC, completeness results
are \emph{relative} because,
by adoption of proof rule~\ref{ax:DE},
valid data identities may be used in derivations. 
The axiom system \BTC\ is complete for closed
(no data variables, no tuplix variables) terms.
In the proof we use canonical terms: 
a \BTC\ canonical term is an alternative composition
\[
  t_1+\cdots+t_k
\]
of \CTC\ canonical terms for some $k\geq 0$
(in case $k=0$, this term is defined as \nt).
Clearly, we can derive such a canonical
term for every \BTC\ term by pushing ${+}$
outward using axiom~\ref{ax:C5}.

\begin{theorem}
  \label{thm:btccomp}
	For closed terms, \BTC\ is complete
	with respect to the standard model, i.e.,
	for closed terms $s$ and $t$,
	$s\sim t$ implies $\BTC\vdash s=t$.
\end{theorem}

\begin{proof}
Take closed terms $s$ and $t$ with $s\sim t$.
We may assume for $s$ and $t$ that
there are respective canonical terms
\[
  s_1+\cdots+s_k\quad\text{and}\quad t_1+\cdots+t_l
\]
such that $\sem{s_i}$ and $\sem{t_j}$
are singleton sets for $i=1,\ldots,k$
and $j=1,\ldots,l$.
Since $\sem{s}=\sem{t}$, it is clear 
that for every $s_i$ there is a
$t_j$ such that $s_i\sim t_j$, and vice versa.
By completeness of \CTC\ these are derivably equal.
Then, $s=t$ can be derived using
axioms \ref{ax:C1}--\ref{ax:C4}.
\end{proof}

Completeness for open terms
(which we did prove for \CTC)
appears to be more involved.
We leave this open for future work.

\section{Zero-Test Logic}
\label{sec:ZTL}
We have seen how the zero-test operator $\ztest{p}$
tests the equality $p=0$.
Using axioms~\ref{ax:T6} and \ref{ax:DE},
it is easy to derive the following identities,
which we shall often use implicitly in derivations:
\begin{align*}
\ztest{u} &= \ztest{-u},\\
\ztest{u} &= \ztest{u/n},\\
\ztest{u} &= \ztest{n\cdot u},
\end{align*}
where $n$ ranges over all non-zero numerals.

We present some observations on the use of the
zero-test operator which lead to a simple logic.

First, the empty tuplix \et\
with $\et=\ztest 0$ by axiom \ref{ax:T7}
may be read as `true', and
the null tuplix \nt\ 
with $\nt=\ztest 1$ by axiom \ref{ax:T8}
may be read as `false'.

Negation.
Define the test `not $p=0$' by
\[
  \nztest{p}\defeq \ztest{1-p/p}.
\]

Conjunctive composition of tests may be read as
logical conjunction:
\[
  \ztest{p}\conjc\ztest{q} 
   \stackrel{\eqref{ax:T9}}{=} \ztest{p/p+q/q}
\]
tests `$p=0$ and $q=0$'.

Alternative composition of tests may be read as
logical disjunction:
\[
  \ztest p+\ztest q\stackrel{\eqref{ax:C6}}{=}\ztest{pq}
\]
tests `$p=0$ or $q=0$'.

A \emph{formula} would then be 
a tuplix-closed (no tuplix variables)
\BTC\ term without entries.
Any formula can be expressed as a single test \ztest{p} 
using axioms \ref{ax:T7}--\ref{ax:T9} and
\ref{ax:C6}, and the definition of negation.
Let $\varphi$ range over formulas, and write 
$\widetilde\varphi$ for the negation of $\varphi$.

We find that this logic has all the usual properties.
Clearly, conjunction and disjunction are
commutative, associative, and idempotent,
and it is not difficult to derive
distributivity, absorption, and
double negation elimination.
The following identities are easily derived as well:
\begin{align}
\label{eq:logic1} \varphi + \widetilde\varphi &= \et,\\
\label{eq:logic2} \varphi \conjc \widetilde\varphi &= \nt,\\
\varphi +\et &= \et,\\
\varphi\conjc\nt &= \nt.
\end{align}
As usual, implication can be defined
in terms of negation and disjunction:
\[
\nztest{p}+\ztest{q} = \ztest{(1-p/p)\cdot q}
\]
tests `$p=0$ implies $q=0$'.

\begin{note}
If the absolute operator $|\_|$
(with $|p|=p$ if $p\geq 0$, and $|p|=-p$ otherwise)
is added to the signature of the data type,
we can also express inequalities:
\[
  \ztest{|q-p|=q-p}
\]
expresses the test $p\leq q$. 
\end{note}

\section{Generalized Alternative Composition}
\label{sec:gensum}
The \emph{generalized alternative composition
(\emph{or:} summation) operator} $\gsum_u$
is a unary operator that 
\emph{binds} data variable $u$ and can be seen
as a data-parametric generalization of 
the alternative composition operator $+$.
We add this binder to the signature of \BTC\
and write $\FV(t)$ for the set of free data 
variables occurring in tuplix term $t$.
We write $\V(p)$ for the set of data 
variables occurring in data term $p$
(there is no variable binding within data terms).
Define substitution $t[p/u]$ as: replace
every free occurrence of data variable $u$
in tuplix term $t$ by the data term $p$,
such that no variables of $p$ become bound
in these replacements.
E.g., recall the proof rule \ref{ax:subst}:
\[
  t\conjc\ztest{u-p} = t[p/u]\conjc\ztest{u-p}.
\]
This rule remains sound in the setting with summation,
but application of the rule may
require the renaming of bound variables in $t$
using axiom \ref{ax:S2}, see below,
so that the substitution can be performed.
When considering substitutions we shall implicitly
assume that bound variables are renamed properly.

The axiom schemes for $\gsum_u$ are
as follows, where $s$ and $t$ range over
tuplix terms and $p$ ranges over data terms.
\begin{align}
\tag{\axname{S1}}\label{ax:S1}
	 \gsum_u t &= t 
	&\text{if }u\not\in\FV(t)\\
\tag{\axname{S2}}\label{ax:S2}
	 \gsum_u t &= \gsum_vt[v/u]
	&\text{if }v\not\in\FV(t)\\
\tag{\axname{S3}}\label{ax:S3}
	 \gsum_u (s\conjc t) &= s\conjc\gsum_u t
	&\text{if }u\not\in\FV(s)\\
\tag{\axname{S4}}\label{ax:S4}
	 \gsum_u (s+t) &= \gsum_u s + \gsum_u t\\
\tag{\axname{S5}}\label{ax:S5}
	 \gsum_u\ztest{u-p} &= \et
	&\text{if }u\not\in\V(p)\\
\tag{\axname{S6}}\label{ax:S6}
	 \gsum_u\nztest{u-p} &= \et
	&\text{if }u\not\in\V(p)
\end{align}
(Recall from Section~\ref{sec:ZTL} that 
\nztest{p} is defined as \ztest{1-p/p}.)

The standard model for \BTC\
is extended with the following
interpretation of summation:
\[
  \sem{\gsum_u t} \defeq
   \{\sem{t[p/u]}~|~p \text{ a closed data term} \}.
\]
The axiom schemes \ref{ax:S1}--\ref{ax:S6}
are sound with respect to this model.

\begin{note}
A similar summation operator
(binding of data variables
that generalizes alternative composition)
is part of the specification
language $\mu$CRL~\cite{GP95},
which combines the process algebra ACP~\cite{BK84}
with equationally specified abstract data types.
A detailed exposition of a semantics and proof theory
for this `choice quantification' in the setting of
process algebra can be found in the work of Luttik~\cite{Lut03}.
A corresponding treatment is possible in our case.
\end{note}

\begin{lemma}
The following identities are derivable
for all data terms $p$ with $u\not\in\V(p)$.
\begin{align}
\label{eq:sumid1}
  \gsum_u (t\conjc\ztest{u-p}) &= t[p/u] \\
\label{eq:sumid2}
  \gsum_u t &= t[p/u] + \gsum_u t\\
\label{eq:sumid3}
  \gsum_u t &= t[p/u] + \gsum_u (t \conjc\nztest{u-p})
\end{align}
\end{lemma}

\begin{proof}
Derivation of~\eqref{eq:sumid1}:
\begin{align*}
\gsum_u (t\conjc\ztest{u-p}) 
  &= \gsum_u (t[p/u]\conjc\ztest{u-p})\\
  &= t[p/u]\conjc\gsum_u \ztest{u-p}\\
  &= t[p/u]
\end{align*}
using \ref{ax:subst}, \ref{ax:S3} and \ref{ax:S5}.
Derivation of \eqref{eq:sumid2}:
\begin{align*}
\gsum_u t &=
  \gsum_u(t\conjc\ztest{u-p})+\gsum_u(t\conjc\nztest{u-p})\\
  &= t[p/u] +\gsum_u t.
\end{align*}
using \ref{ax:T3}, \eqref{eq:logic1}, \ref{ax:S4},
\ref{ax:C3}, and \eqref{eq:sumid1}.
Derivation of \eqref{eq:sumid3}: similar.
\end{proof}

\begin{example}
We derive
\[ 
  \gsum_u(a(u)\conjc\ztest{u^2-1})=a(-1)+a(1).
\]
Proof: from
\[
    \ztest{u^2-1}
  = \ztest{(u+1)(u-1)}
  = \ztest{u+1}+\ztest{u-1}
\]
it follows that
\begin{align*}
\gsum_u(a(u)\conjc\ztest{u^2-1})
  &= \gsum_u(a(u)\conjc\ztest{u+1})+
    \gsum_u (a(u)\conjc\ztest{u-1})\\
  &=a(-1)+a(1)
\end{align*}
using \eqref{eq:sumid1}.
\end{example}

\begin{example}\label{ex:let}
\emph{Let-expressions} or \emph{let-bindings}
allow value declarations or partial bindings in expressions.
The term
\[
  \gsum_u(t\conjc \ztest{u-p})
\]
characterizes 
\[
  \text{let $u=p$ in $t$}.
\]
Of course, $p$ may contain variables, as for instance 
`let $u=7v+1$ in $t$'
can simply be expressed as
\[
  \gsum_u(t\conjc\ztest{u-7v-1}).
\]
\end{example}

\section{Auxiliary Operators}
\label{sec:aux}
For \BTC\ with summation, 
we define three auxiliary operators:
scalar multiplication, clearing, and encapsulation.
In each case the axioms for 
choice and summation (numbered 6 and 7)
can be omitted, for inclusion in
axiom system \CTC\ or \BTC.

\subsection{Scalar Multiplication}
Scalar multiplication $p\cdot t$
multiplies the quantities contained in entries
in tuplix term $t$ by $p$.
It is specified by means of
the following axioms.
\begin{align}
\tag{\axname{Sc1}} u\cdot\et &= \et\\
\tag{\axname{Sc2}} u\cdot\nt &= \nt\\
\tag{\axname{Sc3}} u\cdot\ztest v &= \ztest v\\
\tag{\axname{Sc4}} u\cdot a(v) &= a(u\cdot v)\\
\tag{\axname{Sc5}}
  u\cdot (x\conjc y) &= u\cdot x \conjc u\cdot y\\
\tag{\axname{Sc6}}
  u\cdot (x+y) &= u \cdot x + u \cdot y\\
\tag{\axname{Sc7}}
  p\cdot\gsum_vt &= 
    \gsum_v(p\cdot t) &\text{if }v\not\in\V(p)
\end{align}
Axiom~\axname{Sc7} is an axiom scheme
with $p$ ranging over data terms and $t$ ranging over tuplix terms.
An example with scalar multiplication
is given in Section~\ref{sec:vb1}.

\paragraph{Standard Model.}
Take the standard model
$\SM(\DM,\Attr)$ as before (see Section~\ref{sec:SM}).
For partial function $f\in\PF$ and value $d\in\DM$, define
the scalar multiplication $d\cdot f$ as expected:
$(d\cdot f)(a) = d \cdot (f(a))$ if $f(a)$ is defined, and 
undefined otherwise.
The interpretation of scalar multiplication is defined by
\[
  \sem{p\cdot t} \defeq \{ \sem p\cdot f~|~f\in\sem{t}\}.
\]

\subsection{Clearing}\label{sec:clear}
For set of attributes $I\subseteq \Attr$, the operator
\[
  \Clr_I(x)
\]
renames all entries of $x$ with attribute in $I$ to \et.
It ``clears'' the attributes contained in $I$.
Axioms:
\begin{align}
\tag{\axname{Cl1}}\Clr_I(\et) &= \et\\
\tag{\axname{Cl2}}\Clr_I(\nt) &= \nt\\
\tag{\axname{Cl3}}\Clr_I(\ztest u) &= \ztest u\\
\tag{\axname{Cl4}}\Clr_I(a(u)) &=
  \begin{cases}
   \et  & \text{if }a\in I\\
   a(u) & \text{otherwise}
  \end{cases}\\
\tag{\axname{Cl5}}
  \Clr_I(x\conjc y) &=\Clr_I(x)\conjc\Clr_I(y)\\
\tag{\axname{Cl6}}
  \Clr_I(x+y) &= \Clr_I(x) + \Clr_I(y)\\
\tag{\axname{Cl7}}
  \Clr_I(\gsum_ut) &=\gsum_u(\Clr_I(t))
\end{align}

For a set of attributes $B\subseteq\Attr$
one can think of a function
\[
  \mathit{Select}_B(x)\defeq \Clr_{\Attr\setminus B}(x).
\]
This function allows to focus on those entries with
attribute contained in $B$.

\paragraph{Standard Model.}
Take the standard model $\SM(\DM,\Attr)$ as before
(see Section~\ref{sec:SM}).
Define the function $\Clr_I$ on elements of \PF\ as follows.
For partial function $f\in\PF$ and attribute $a\in\Attr$, 
if $f(a)$ is undefined or $a\in I$,
then $\Clr_I(f)(a)$ is undefined,
else $\Clr_I(f)(a)=f(a)$.
The interpretation of clearing:
\[
  \sem{\Clr_I(t)} \defeq \{ \Clr_I(f)~|~f\in\sem t\}.
\]

\subsection{Encapsulation}
Encapsulation can be seen as `conditional clearing'.
For set of attributes $H\subseteq\Attr$,
the operator $\Enc_H(x)$ encapsulates all entries in $x$
with attribute $a\in H$.
That is, for $a\in H$, if the accumulation of quantities
in entries with attribute $a$ equals zero,
the encapsulation on $a$ is considered successful
and the $a$-entries are \emph{cleared\/} (become \et);
if the accumulation is not equal to zero, 
they become null (\nt).
This accumulation of quantities is
computed per alternative:
the encapsulation operator distributes over alternative
composition.
Axioms:
\begin{align}
\tag{\axname{E1}}\label{ax:E1}
  \Enc_H(\et) &= \et\\
\tag{\axname{E2}}\label{ax:E2}
  \Enc_H(\nt) &= \nt\\
\tag{\axname{E3}}\label{ax:E3}
  \Enc_H(\ztest{u}) &= \ztest{u}\\\displaybreak[0]
\tag{\axname{E4}}\label{ax:E4}
  \Enc_H(a(u)) &=
  \begin{cases}
   \ztest{u} & \text{if } a\in H\\
   a(u)      & \text{if } a\not\in H
   \end{cases}\\\displaybreak[0]
\tag{\axname{E5}}\label{ax:E5}
  \Enc_H(x\conjc\Enc_H(y)) &=
  \Enc_H(x)\conjc\Enc_H(y)\\
\tag{\axname{E6}}\label{ax:E6}
  \Enc_H(x+y)&= \Enc_H(x)+\Enc_H(y)\\
\tag{\axname{E7}}\label{ax:E7}
  \Enc_H(\gsum_ut) &=\gsum_u(\Enc_H(t))
\end{align}
We further define
\[
  \Enc_{H\cup H'}(x)\defeq\Enc_H\circ\Enc_{H'}(x).
\]

\paragraph{Standard Model.}
Take the standard model $\SM(\DM,\Attr)$ as before.
We say that a partial function $f$ in \PF\
is \emph{neutral} on attribute $a$,
if either $f(a)$ is undefined or $f(a)=0$.
We interpret encapsulation as follows.
\[
  \sem{\Enc_H(t)} \defeq
   \{ \Clr_H(f)~|~
        f\in\sem t,\
        f \text{ neutral on all }a\in H
   \},
\]
where $\Clr_H$ is as defined in Section~\ref{sec:clear}.

\subsection{On the Derivation of Encapsulations}
By the derivation of an encapsulation we mean the
elimination of the encapsulation operator
by application of its defining axioms from
left to right. Of course,
this elimination is in general only possible
for tuplix-closed terms.
We present some helpful identities and
example derivations.

We start with a lemma.

\begin{lemma}\label{lem:E1}
For all tuplix-closed terms $t$,
if no element of set of attributes $H$ occurs in $t$,
then
\[
  \Enc_{H}(t)	= t
\]
and, for any term $s$,
\[
  \Enc_{H}(s\conjc t) 
	=
	\Enc_{H}(s) \conjc t
\]
are derivable.
\end{lemma}
\begin{proof}
The first identity is easy,
using structural induction on term $t$.
Then, the second one follows using axiom~\ref{ax:E5}:
\[
  \Enc_{H}(s\conjc t) = \Enc_{H}(s\conjc \Enc_{H}(t))
	=
	\Enc_{H}(s) \conjc \Enc_{H}(t)
	=
	\Enc_{H}(s) \conjc t.
\]
\end{proof}

When deriving an encapsulation,
we generally
split the encapsulation up:
for $a\in H$, we have by definition that
\[
  \Enc_H(t)=\Enc_{H\setminus\{a\}}\circ\Enc_{\{a\}}(t),
\]
and we start with $\Enc_{\{a\}}(t)$.
Observe that encapsulation distributes
over (generalized) alternative composition,
so we can push it inward until we reach an
conjunctive composition
in which we assume that the $a$-entries
have been accumulated into a single entry
using axiom~\ref{ax:T5}.
So this yields an application of the form
\[ \Enc_{\{a\}}(a(p)\conjc t')
\]
where $a$ does not occur in $t'$, so using Lemma~\ref{lem:E1},
this is equal to
\[ \ztest{p}\conjc t'.
\]

Example:
\begin{align*}
\Enc_{\{b\}}(a(-3)\conjc b(1)\conjc b(2)\conjc b(-3)
\conjc c(3))
  &=\Enc_{\{b\}}(b(0)\conjc a(-3)\conjc c(3))\\
  &=\Enc_{\{b\}}(b(0))\conjc a(-3)\conjc c(3)\\
  &=\ztest{0}\conjc a(-3)\conjc c(3)\\
  &=a(-3)\conjc c(3).
\end{align*}
Another example:
\[ \Enc_{\{a,b\}}(a(0)\conjc b(0)) 
   = \Enc_{\{a\}}\circ\Enc_{\{b\}}(a(0)\conjc b(0))
   = \Enc_{\{a\}}(a(0))
   = \et.
\]

In applications that use
information hiding (summation),
we typically encounter encapsulations
like this one:
\[ \Enc_{\{a\}}(a(2)\conjc 
   \gsum_u(a(-u)\conjc b(u/2)\conjc c(u/2))) = 
   b(1)\conjc c(1),
\]
where instantiation of the hidden variable $u$
is enforced by the encapsulation.

Let's see how to derive such encapsulations.
First, we 
have, for data term $p$ with $u\not\in\V(p)$, and
tuplix-closed term $t$ that does not contain $a$,
\[ a(p)\conjc\gsum_u(a(q)\conjc t) = 
  \gsum_u(a(p+q)\conjc t).
\]
Then we easily find that
\[ \Enc_{\{a\}}(a(p)\conjc\gsum_u(a(q)\conjc t)) = 
  \gsum_u(\ztest{p+q}\conjc t)
\]
using Lemma~\ref{lem:E1}.
In the particular case that $q=-u$ we
find
\[
  \Enc_{\{a\}}(a(p) \conjc\gsum_u(a(-u)\conjc t))
	=
	t[p/u]
\]
using \eqref{eq:sumid1}.
Another example:
\begin{align*}
\Enc_{\{a\}}(a(-6)\conjc\gsum_u(a(2u)\conjc t)) 
 &= 
  \gsum_u(\ztest{2u-6}\conjc t)\\
 &=\gsum_u(\ztest{u-3}\conjc t)\\
 &= t[3/u].
\end{align*}

In the next example, the
instantiation is determined
within the summation:
\begin{align*}
\Enc_{\{a\}}(\gsum_u(a(u+1)\conjc b(-u/2)))
&=
\gsum_u(\ztest{u+1}\conjc b(-u/2))
\\
&=
b(1/2).
\end{align*}

In a similar way, one can reduce
\[
\Enc_{\{a\}}(
  \gsum_u(
    a(-u) \conjc b(u/2) \conjc c(-u/2)
    \conjc a(200) \conjc b(-50) \conjc c(-150)
  )
)
\]
to
\[
  b(50)\conjc c(-250).
\]

\section{Example: Incremental Budgeting}
\label{sec:vb1}

A financial budget is modeled as a tuplix.
We let an entry $a(p)$ represent a payment: 
the attribute $a$ is used in the communication
between payer and payee,
and describes or identifies a transaction;
we also refer to the attribute as the \emph{channel}
of the transaction, and say that
the payment occurs \emph{along} the channel. 
The term $p$ represents the amount of money involved.
An entry $a(p)$ with $p>0$ stands for
an obligation to pay amount $p$ along channel $a$.
If $p<0$, the entry stands for the expected receipt
of amount $p$ along $a$.

In the following example we consider some annual budgets.
In order to simplify descriptions it is assumed
that various payments are due twice
per year only, during periods \perA\ and \perB.
Attributes of the form $a_\perA$ and $a_\perB$ are used
in the specification of payments during these respective
periods.
Examples with monthly, weekly or daily payments can be
given in a similar fashion. 

Consider a budget $\Budget_{2006}$
containing the financial results
of some entity in year 2006.
E.g., take
\[
\Budget_{2006} =
  a_{\perA}(30) \conjc
  a_{\perB}(30) \conjc
  b_{\perA}(20) \conjc
  b_{\perB}(25) \conjc
  c_{\perA}(-107).
\]
On the basis of this realized budget, an
allocated budget for 2007 covering corresponding entries
could be specified as, e.g.,
\[
\Budget_{2007} =
  a_{\perA}(32) \conjc
  a_{\perB}(32) \conjc 
  b_{\perA}(21) \conjc
  b_{\perB}(28) \conjc
  c_{\perA}(-116).
\]

Assuming that a 2008 budget is to be
determined without having 2007
realization figures available,
several ways to adapt the 2007 budget to a
2008 budget can be imagined.
The widespread (and well-documented, see, e.g.,~\cite{Tuc82})
strategy of \emph{incremental budgeting}
implies that the 2007 budget is taken as the
point of departure for designing a 2008 budget.
For 2008 one may consider two possible budgets:
an ad hoc increase
of each entry, leading to something like
\[
\Budget_{2008} =
  a_{\perA}(33) \conjc
  a_{\perB}(33) \conjc
  b_{\perA}(22) \conjc
  b_{\perB}(30) \conjc
  c_{\perA}(-123),
\]
or, alternatively, 
\[
  \Budget'_{2008} = (1 + (i/100)) \cdot \Budget_{2007}
\]
which adjusts each 2007 entry
with the same inflation percentage $i$.

Yet another option for a 2008 budget is to adjust
the 2006 realization $\Budget_{2006}$ with
inflation twice. This yields
\[
  \Budget''_{2008} = (1 + (i/100))^2 \cdot \Budget_{2006}.
\]

Still another option for the definition of a 2008
budget is the average
\[
  \Budget'''_{2008} =
  	(1/2)\cdot 
  	(\Budget'_{2008} \conjc \Budget''_{2008})
\]
of the latter two budgets.

\section{Example: Modular Budget Design}
\label{sec:vb2}
Modular financial budget design is a necessity in 
large organizations, assuming that budgets
are at all used, i.e., that `beyond budgeting'~\cite{bb}
is not (yet) the dominant strategy for financial planning.
Financial budgets are probably the most complex budgets
around which calls for modularity. 
Surprisingly, however, we have not been able to find
any literature about the subject of formalized modular
budget design.
In the example of this section we will outline
how tuplix calculus can support modular budget
descriptions in a meaningful way.
The example is presented in abstract terms but
its origin is practical.

We consider an organization that 
consists of the following constituents:\footnote{%
	In the practical case behind the example,
	\source\ is a university division,
	\control\ represents a graduate school, 
	the \pu s represent different master programs, 
	\service\ provides various forms of support ranging
	from student counseling to timetabling,
	and \capa\ represents a department from which
	educational staff will be used.
} 
\begin{itemize}
\item Part \source\ is a \emph{financial source}
  which correlates with production figures.
\item Part \control\
  is a \emph{control group} that dispatches 
  the incoming financial stream to the 
  production units and the service center \service.
\item Parts \pua\ and \pub\ are 
  \emph{production units}. 
  These units produce
  the same two types of products
  (type 1 and type 2).
\item Part \service\ is a shared
  \emph{service center}
  providing services needed by the 
  production units.
\item Part \capa\ is a \emph{capacity group} from which
  both production units draw manpower. 
\end{itemize}

Streams of money between these parts run as depicted in
this figure:
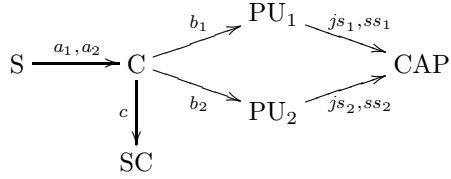
\[
\xymatrix@R=5pt@C=30pt{
 && \pua\ar[dr]^{\attrname{js}_1,\attrname{ss}_1} &\\
 \source \ar[r]^{\attrname{a}_1,\attrname{a}_2}&
   \control\ar[dd]_{\attrname{c}}
   \ar[ur]^{\attrname{b}_1}\ar[dr]_{\attrname{b}_2} && \capa\\
 && \pub\ar[ur]_{\attrname{js}_2, \attrname{ss}_2} &\\
 &\service
}
\]
The labels ($\attrname{a}_1$, $\attrname{a}_2$, etc.) 
on the arrows in this picture
are attribute names that will be used
in the specification of payments.

The financial source rewards the production by the
production units: for each product that is produced,
a constant reward (depending on the type of the product)
is allocated to the control group \control.
The control group will dispatch the rewards
for both product types to the production units
and to the service center.
The production units receive money from \control,
and pay money to the capacity group in return of
junior staff capacity as well as senior staff capacity.

We specify budgets for the parts
\source, \control, \pua\ and \pub,
and we will examine how to compose
one joint budget $B$ from these.
All budgets involved specify the same period of
time (e.g., the calendar year 2008).
We take a stepwise approach and specify
the budgets in two phases taking 
increasingly more aspects into account. 
In the first stage, 
both production units obtain an equal
reward, independent of their
contribution to the total production.
The budgets are defined as follows.
\begin{itemize}
\item The financial source \source\
rewards production:
for each product of type $i$ that is produced
($i=1,2$),
a constant reward $\reward_{i}$ is allocated to
the control unit \control.
For production unit $\pu_i$ and
product type $j$,
the data variable
\[ \nop_{ij}
\]
stands for the {number of products} of type $j$
produced by $\pu_i$ during the period that is covered.

The budget:
\[ \Budget_{\source} \defeq
    \attrname{a}_1(\Rone\cdot(\nop_{11}+\nop_{21}))
    \conjc
    \attrname{a}_2(\Rtwo\cdot(\nop_{12}+\nop_{22}))
\]
\item The control unit \control\ 
receives the rewards from the financial source.
The amount paid to the service center
\service\
is a fraction \sfrac\ (a value between 0 and 1) 
of the incoming stream,
independently of the use that is made of it.
It further pays each production unit
half of the reward total.
(Observe that, unless $k=0$,
this budget is not \emph{balanced}:
the expenses are higher than the income.)
\begin{align*}
 \Budget_{\control}\defeq{}&
   \gsum_{u}\gsum_v (\\
& \quad
    \attrname{a}_1(-u)\conjc
    \attrname{a}_2(-v)\conjc{}\\
& \quad
    \attrname{c}(\sfrac\cdot (u+v))\conjc{}\\
& \quad
    \attrname{b}_1((u+v)/2)\conjc
    \attrname{b}_2((u+v)/2)   )
\end{align*}

\item Budgets for the production units:
\begin{align*}
\Budget_{\pua} &\defeq
   \gsum_{u}
   (\attrname{b}_1(-u)\conjc
    \attrname{js}_1(u/2)\conjc
    \attrname{ss}_1(u/2)
   )\\
\Budget_{\pub} &\defeq
   \gsum_{u}
   (\attrname{b}_2(-u)\conjc
    \attrname{js}_2(u/2)\conjc
    \attrname{ss}_2(u/2)
   )
\end{align*}
In this first version the production units
obtain an equal amount of funding,
which is spent in equal parts on 
senior staff (via $\attrname{ss}_1$ and $\attrname{ss}_2$)
and on junior staff
(via $\attrname{js}_1$ and $\attrname{js}_2$).
\end{itemize}
The combined budget:
\[ \Budget\defeq 
   \Enc_{\{ \attrname{a}_1,\attrname{a}_2,\attrname{b}_1,
         \attrname{b}_2\}}
   (\Budget_{\source}\conjc \Budget_{\control}\conjc 
    \Budget_{\pua} \conjc \Budget_{\pub}).
\]

We find
\begin{align*}
\Budget ={}&
  \gsum_u ( \\
& \quad
   \ztest{u - 
	 \Rone\cdot(\nop_{11}+\nop_{21}) -
	 \Rtwo\cdot(\nop_{12}+\nop_{22})}\conjc{}\\
& \quad
  \attrname{c}(\sfrac\cdot u)\conjc
  \attrname{js}_1(u/4)\conjc
  \attrname{ss}_1(u/4)\conjc
  \attrname{js}_2(u/4)\conjc
  \attrname{ss}_2(u/4)  ).
\end{align*}
A straightforward derivation of this identity
leads to a closed term without summation;
in the expression above
we have introduced a `let-binding'
(see the example in Section~\ref{sec:gensum})
with variable $u$ to improve the readability.

In the second stage of the budget,
we take into account that the production units
need funding proportional to their production volume,
and may spend their resources on senior staff capacity
and junior staff capacity in different proportions.
Moreover, the control unit also charges
the production units for the
costs of the services provided by \service.
This leads to the following refinement
of the budgets:
\begin{align*}
 \Budget_{\control}\defeq{}&
   \gsum_{u}\gsum_v (\\
& \quad
    \attrname{a}_1(-u)\conjc
    \attrname{a}_2(-v)\conjc{}\\
& \quad
    \sfrac\cdot \attrname{c}(u+v)\conjc{}\\
& \quad
   (1-\sfrac)\cdot
    (\attrname{b}_1(
      (\nop_{11}/(\nop_{11}+\nop_{21}))u+
      (\nop_{12}/(\nop_{12}+\nop_{22}))v)\conjc{}\\
& \quad\phantom{(1-\sfrac)\cdot (}
     \attrname{b}_2(
      (\nop_{21}/(\nop_{11}+\nop_{21}))u+
      (\nop_{22}/(\nop_{12}+\nop_{22}))v)
   ))
\\\displaybreak[0]
 \Budget_{\pua} \defeq{}&
   \gsum_{u}
   (\attrname{b}_1(-u)\conjc
    \attrname{js}_1((1/4)u)\conjc
    \attrname{ss}_1((3/4)u)
   )
\\
\Budget_{\pub} \defeq{}&
   \gsum_{u}
   (\attrname{b}_2(-u)\conjc
    \attrname{js}_2((1/3)u)\conjc
    \attrname{ss}_2((2/3)u)
   )
\end{align*}

Additional phases that take more aspects into
account can be easily imagined.
For instance both production units
may be given a fixed amount of funding
independent of production volume and the remaining funding
spread in proportion with production volume.
That distribution strategy for \control\
allows one unit to proceed when its production
is low thus awaiting a next phase with better circumstances.

\section{Conclusion}
\label{sec:concl}
We have introduced a calculus for tuplices.
It has an underlying data type called quantities
which is required to be modeled
by a zero-totalized field.
We started with the core tuplix calculus \CTC\
for entries and tests, which are
combined using conjunctive composition.
We defined a standard model and proved
that \CTC\ is relatively complete with respect to it.
We further defined operators for
choice, information hiding,
scalar multiplication, clearing and encapsulation.
We ended with two examples of applications;
one on incremental financial budgeting, and one
on modular financial budget design.

We refer to~\cite{TFB}
for a discussion on the formalization
of financial budgets.
It also contains a more elaborate
application of the tuplix calculus
in the style of the example in Section~\ref{sec:vb2}.

Further related work seems to be scarce.
We mention here the work of Elsas et al.~\cite{Els96,GER}
on audit theory, and the work of
Bergstra and Middelburg~\cite{BM} on 
computational capital.
Both are theoretical approaches that apply process theory
in the analysis of organizations dealing with money streams:
the former uses Petri nets,
the latter process algebra.
In this, they focus more on behavioral aspects
than we do.

An immediate issue for future work is
the completeness of \BTC\ for open terms,
and consequently the completeness of \BTC\ with summation.
We would further like to connect this
theory to the formalization of interface
groups and financial transfer architectures
studied in~\cite{BP2007}.

\end{document}